\documentclass[sigconf, nonacm]{acmart}
\AtBeginDocument{%
  \providecommand\BibTeX{{%
    \normalfont B\kern-0.5em{\scshape i\kern-0.25em b}\kern-0.8em\TeX}}}
\usepackage{amsmath}
\DeclareMathOperator*{\argmax}{argmax} % thin space, limits underneath in displays
\usepackage{algorithm}
\usepackage{algpseudocode}
\usepackage {tikz}
\usetikzlibrary {positioning}
\usepackage{multirow}
\usepackage{caption}
\usepackage{diagbox}
\usepackage{graphicx}
\usepackage{textcomp}
\usepackage{xcolor}
\usepackage{calrsfs}
\usepackage{array}
\usepackage{caption}
\newcolumntype{C}[1]{>{\centering\arraybackslash}m{#1}}
\begin{document}

\title{Variational Inference for Category Recommendation in E-Commerce platforms  }

\author{Ramasubramanian Balasubramanian}
\authornote{Both authors contributed equally to this research.}
\email{r.balasubramanian@walmart.com}
\orcid{0000-0001-5343-4636}
\affiliation{%
  \institution{Walmart Global Tech}
  \streetaddress{600 W California Ave}
  \city{Sunnyvale}
  \state{California}
  \country{USA}
  \postcode{43017-6221}
}

\author{Venugopal Mani}
\authornotemark[1]
\email{venugopal.mani@walmart.com}
\affiliation{%
  \institution{Walmart Global Tech}
  \streetaddress{600 W California Ave}
  \city{Sunnyvale}
  \state{California}
  \country{USA}
  \postcode{43017-6221}
}

\author{Abhinav Mathur}
\email{amathur1@walmart.com}
\affiliation{%
  \institution{Walmart Global Tech}
  \streetaddress{600 W California Ave}
  \city{Sunnyvale}
  \state{California}
  \country{USA}
  \postcode{43017-6221}
}

\author{Sushant Kumar}
\email{sushant.kumar@walmart.com}
\affiliation{%
  \institution{Walmart Global Tech}
  \streetaddress{600 W California Ave}
  \city{Sunnyvale}
  \state{California}
  \country{USA}
  \postcode{43017-6221}
}

\author{Kannan Achan}
\email{kannan.achan@walmartlabs.com}
\affiliation{%
  \institution{Walmart Global Tech}
  \streetaddress{600 W California Ave}
  \city{Sunnyvale}
  \state{California}
  \country{USA}
  \postcode{43017-6221}
}

\renewcommand{\shortauthors}{Balasubramanian and Mani, et al.}

\begin{abstract}
Category recommendation for users on an e-Commerce platform is an important task as it dictates the flow of traffic through the website. It is therefore important to surface precise and diverse category recommendations to aid the users' journey through the platform and to help them discover new groups of items. An often understated part in category recommendation is users' proclivity to repeat purchases. The structure of this temporal behavior can be harvested for better category recommendations and in this work, we attempt to harness this through variational inference. Further, to enhance the variational inference based optimization, we initialize the optimizer at better starting points through the well known Metapath2Vec algorithm. We demonstrate our results on two real-world datasets and show that our model outperforms standard baseline methods.
\end{abstract}





\maketitle

\section{Introduction}
Personalizing user experience drives both user satisfaction and key business goals such as engagement, conversion, revenue, etc. for e-Commerce websites. Recommender systems are a common way to personalize e-Commerce journeys. They predict user behavior and make the user journey easier. Among recommender systems in large scale e-Commerce platforms, the category recommendation task is an important one as it performs twin services of guiding a user to parts of the website that they are likely to be interested in as well as to categories that have not been explored much by the user, thus helping in boosting the platform's revenue. 

To achieve category recommendation, a collaborative filtering framework has traditionally worked well \cite{ji_category_2014}. In e-Commerce, explicit signals like ratings or reviews are very sparse and the main sources of signals are the implicit actions of the users. These implicit actions like views, transactions of an item, etc. can in turn be leveraged to form abstract representations of the users. These abstract representations are often in the form of \emph{embeddings} which can be obtained in a variety of different ways depending on the way the feature space is formed from the implicit signals. 

Users in e-Commerce settings (especially grocery) display repeat purchase patterns and there are two unique structures in this implicit interaction data: the graphical structure and the temporal structure. The graphical structure is a natural way to represent e-Commerce data, often done in the form of a \emph{heterogeneous network} \cite{dong_m2v_2017}, formed as a graph from user interaction data. This interaction graph can then be used to develop rich embeddings for a variety of tasks including that of node similarities and correlations. Node correlations are in turn used for recommendations. Using interaction graphs to form embeddings has been achieved through a variety of algorithms like Item2Vec \cite{barkan_i2v_2016}, Prod2Vec \cite{grbovic_p2v_2015}, and Metapath2Vec \cite{wan_t2v_2018}. 

However, in these graph based models, the temporal structure of the data is not focused upon. In domains like grocery, temporal structures play a very important role in user interactions. For example, a user is likely to display affinity towards a category like paper towels (buying it every other few months). While the association of the user to the paper towel might be picked up pretty easily, the pattern of \emph{when} the user is likely to purchase can also be harvested for future predictions. It is therefore important to capture both these underlying structures in one framework and we attempt to do this in the collaborative filtering setting by constructing a temporal signal from the implicit data.

We aim to exploit both the semantic relationship between nodes through interaction graph based models as well as the temporal structure of the data through a variational inference based framework. We introduce an additional signal for temporal affinity of users to categories and then use this signal, in addition to the implicit signal, to learn the embeddings. This idea of treating both matrices as observed data, is then used to learn embeddings (latent factors) which explain the graphical and temporal structure of the observed data well. In past works, domain expertise has been used to initialize these latent factors \cite{lavee_satisfaction_2019}. Rather than such an initialization, the graphical structure of the interaction graph can be utilized to initialize the latent factors at values which already have some semantics associated with them. In this work,  we use the semantic representations learned through the graph based model (Metapath2Vec) to initialize the variables in our matrix factorization framework. We then use the additional temporal affinity matrix introduced by us to enhance the semantic embeddings by applying variational inference. 

The rest of the work is organized as follows: Section 2 describes related work in literature, Section 3 proposes and explains our model, Section 4 describes the datasets we used, the baselines we compared against, and the implementation details, Section 5 demonstrates our results on two real world datasets and delves into a qualitative breakdown of our results, and Section 6 sums up the work and points to potential future directions.

\section{Related Work}
Category-based content recommendation is a well-studied problem in literature. Works such as \cite{albadvi_hybrid_2009} deal with  providing  personalized recommendations in the context of online retail stores by extracting user preferences in each product category separately. There is also research around recommendation algorithms based on the category correlations to a user to provide relevant search results in the Web \cite{choi_correlations_2010}, preferred next category prediction using a third-rank tensor optimized by a Listwise Bayesian Personalized Ranking (LBPR) approach \cite{he_poi_2017}, etc. Item category information is leveraged in works such as \cite{wei_category_2012} to overcome the limitations of data sparsity and inaccurate similarity in personalized recommendation systems. Our work develops scalable user representations that capture users' category affinities well and thus can be used as an input to works where category experts are employed to leverage the trade-off between performance and accuracy \cite{hwang_experts_2016} and address the shortcomings of the neighborhood-based methods for recommendations.\par 

Our idea of using temporal affinity to a product category in addition to the binary transaction matrix was inspired by works such as \cite{lavee_satisfaction_2019}, which proposes a multi-task probabilistic matrix factorization model with a dual task objective to predict the appeal of items at the selling stage as well as some satisfaction measure in the long run. Our experience in the domain of e-Commerce and works that deal with modeling repeat purchase recommendations \cite{bhagat_repeat_2018} led us to look at temporal category affinity as our second matrix.\par

Works that exploit the graphical structure of the user-item interaction data in e-Commerce settings involve works like \cite{wan_t2v_2018}, which introduces a new representation learning approach to leverage complementary items, user-item compatibility, and user loyalty. Other works in this domain include those that use knowledge-aware learning with dual product embedding to detect complementary product relationships from noisy and sparse user purchase activities \cite{xu_knowledge_2020}, etc. The Metapath2Vec algorithm used in our work was introduced by Dong et al. in \cite{dong_m2v_2017},  where they formalize metapath based random walks to construct the heterogeneous neighborhood of a node and then leverage a heterogeneous skip-gram model to perform node embeddings. But, maximizing the co-occurrence likelihood with skip-gram-based model formulations limits the expressive ability of the embeddings and the resulting recommendation performance. 

Thus, in our work, we utilize this as the initialization strategy for the variational inference based optimization as these scalable Metapath2Vec representations enable better and faster convergence. We thus leverage information encoded both in the graphical structure as well as the temporal structure of the user purchase data. In terms of the application of Bayesian inference for  recommendation, works like \cite{meng_variational_2019}, where the authors leverage basket context information to develop a context-aware model for grocery recommendation, and \cite{zhou_sbr_2019}, where the authors employ Bayesian inference for parameter estimation of sequential recommendations to develop a session-based recommendation framework, exist in literature.

\section{System Overview}

\subsection{The Category Recommendation Problem}
The Category Recommendation problem is derived from the classic item recommendation task. It may be thought of as a less granular view of the item recommendation task as each user is recommended the top categories that they might be interested in. While it may initially appear intuitive to just recommend at an item level, recommending at a category level has two aspects to it which makes it an interesting problem to solve. The engagement of users with items is a rather sparse signal and the amount of data needed to form meaningful user representations is large. We can develop better user embeddings from the user-category interaction data, as categories capture broader interests as compared to items. These can then be leveraged in a variety of use cases like personalizing the search results for users, providing category-based filters, etc.  The other aspect of category recommendation is that it can lead to the user being exposed to a larger part of the platform as it leads the user to a large group of items rather than individual items. It can serve as a proxy cold-start resolution strategy to help drive users to a subgroup of items that the user is likely to be interested in (due to their past interaction patterns) while also exposing them to new items of that category that a normal item recommendation task would not. 

To formally define the category recommendation problem, consider a set of users in an e-Commerce setting $U = \{u_1, u_2 \dots u_m\}$ and a catalog of items $I = \{i_1, i_2 \dots i_p\}$ that the platform has to offer. Each of these items belong to a certain category from a semantic structure. The categories can be represented as the set $C = \{ c_1, c_2 \dots c_n \} $ such that every item in $I$ has a corresponding mapping to a category in $C$ and $n \ll p$. The category recommendation task then becomes to predict the next category that the user is likely to engage with. Since the value of $n$ (the number of categories) is much lesser than the number of items $p$, the user interaction signal is a much denser one than the sparse user-item relation. The goal of category recommendation is to, for each user $u \in U$, find the top $k$ categories $c_1, c_2 \dots c_k$ that the user is most likely to engage with.

 \subsection{Initialization using Graphical Structure}
Complex real world information is often represented in the form of graphs and for e-Commerce data, a heterogeneous user interaction graph is often a good representation. Like in \cite{dong_m2v_2017}, our heterogeneous graph takes the form of $G = (V,E,T)$ where $V$ is the set of vertices or nodes, $E$ is the set of edges, and $T$ is the set of vertex types. The goal of this graph is to derive structural information and the types of nodes chosen here (the set $T$) are Users, Baskets, and Categories. As in \cite{wan_t2v_2018}, we aim to achieve semantic representations from the \emph{contexts} of a given node (categories or users) to provide good initialization points for the matrix factorization over the temporal signals in the subsequent steps. 

Once the graph structure was set, random walks of different lengths are carried out to define the context of the given node, $C_t(v)$. $C_t(v)$ represents node $v$'s neighborhood with the $t^{th}$ type of nodes. The idea is to integrate a skip-gram model over the interaction graph, known as the heterogeneous skip-gram model, which aims to maximize the heterogeneous context for all the nodes in the graph as per the following equation. 
\begin {equation}\argmax_{\theta} \sum_{v \in V}\sum_{t \in T}  \sum_{c_t \in C_t(v)} \log p(c_t|v;\theta)  \end{equation}

The probability $p(c_t|v;\theta)$, parameterized by the family of parameters $\theta$, is a softmax over the nodes from the context and is equal to $\dfrac{e^{X_{c_t}\cdot X_v}}{\sum_{u \in V} e^{X_u \cdot X_v}}$, where any vector $X_i$ represents the embedding vector of a node $i$ belonging to the graph. Thus, for a given node $v$, the numerator would represent the inner product between the context node and the given node while the denominator would provide normalization over all the nodes in $V$. The popular technique of negative sampling \cite{mikolov_negsamp_2013} is used to make the computation efficient.

The most important step of Metapath2Vec is the transition probability algorithm used to traverse the heterogeneous graph to find the node contexts. Given a node $v$, we need to traverse the heterogeneous graph in a way so that the generated node contexts have semantics. For Metapath2Vec, this is done using a random walk which follows a schema, known as a metapath, to transition over the nodes. In our experiments, we adopt a metapath schema of User-Basket-Category-Basket-User to generate node contexts.  The transition probability between the nodes needs to be carefully defined in order to avoid nodes which are heavily connected from being favored always during the random walks. We adopt a two pronged approach to this similar to an explore-exploit strategy where the transition depends on both the connectivity of a given node while also mutating with a decaying probability to keep the node pool fresh. We experimented with the schema with walks of different lengths (signifying different degrees of granularity). Once the walks are done, the contexts of each of the nodes are generated. The heterogeneous skip-gram model is run over these to generate the embeddings for the users and the categories, which are then used as initialization points for variational inference. 
%\subsection{The Prediction Task}
\subsection{Variational Inference}

This subsection briefly goes over a few different concepts and modeling choices. Wherever possible, we refer the reader to the appropriate resources, for brevity.\par 
In addition to the binary transaction matrix T, we capture a user's dynamically changing affinity to a product category by introducing the temporal affinity matrix A, whose entries capture the number of times a user purchased an item from a given product category, weighted by an exponential decay based on the time elapsed since the purchase \cite{mani_variational_2020}.\par   
We chose to model the values in the binary transaction matrix T and the continuous-valued temporal affinity matrix A with Bernoulli and normal distributions, respectively. We then applied the Jaakkola-Jordan logistic bound \cite{jaakkola_bound_2001} to replace the Bernoulli terms in the likelihood with squared exponentials, as has been done in works like \cite{lavee_satisfaction_2019}. In equation \ref{eq_posterior}, $\theta$ represents the latent variables - the user and category embeddings and biases, H represents the set of all hyperparameters, $u_p$ and $v_q$ are the embedding vectors of the $p^{th}$ user and the $q^{th}$ category, $bu_p$ and $bv_q$ are the corresponding bias vectors, $\alpha$ is a hyperparameter that controls the variance of the corresponding distributions, and $\kappa$ and $\psi$ are the scale and location parameters, which allow the two distributions, $T_{pq}$ and $A_{pq}$, some degree of freedom despite sharing parameters. We used the Metapath2Vec initializations for the embeddings and modeled the other prior variables with normal distributions. The reason for that was that the normal family of distributions is  self-conjugate, thus giving us closed form expressions for the posterior. Hyperparameter tuning and other details \cite{lavee_satisfaction_2019, mani_variational_2020} are omitted here for brevity. 

\begin{equation}\label{eq_posterior}
\begin{split}
P(\theta | T,& A,H) \propto P(T,A|\theta,H)P(\theta|H) \\
&\hspace{-11mm}= \prod_{(p,q,T_{pq})\in T}  P(T_{pq}|u_p, v_q, bu_p, bv_q, \kappa_t, \psi_t, H) \\  &\hspace{-8mm}  \prod_{(p,q,A_{pq})\in A}  P(A_{pq}|u_p, v_q, bu_p, bv_q, \kappa_a, \psi_a, H) \\
&\hspace{-3mm}  \prod_{p = 1}^{m} [P(u_p) P(bu_p|\alpha_{bu})]
\prod_{q = 1}^{n} [P(v_q) P(bv_q|\alpha_{bv})] \\
&\hspace{-3mm} P(\kappa_t|\alpha_{\kappa_t}) P(\psi_t|\alpha_{\psi_t}) P(\kappa_a|\alpha_{\kappa_a}) P(\psi_a|\alpha_{\psi_a})\\
\end{split}
\end{equation}

We could apply Bayes' theorem directly to solve for the posterior seen in equation \ref{eq_posterior}, but that would involve dealing with an intractable high-dimensional integration in the denominator. Thus, we carried out a scalable, approximate Bayesian inference through variational inference. The objective is to find a proxy posterior that is least divergent in terms of the KL (Kullback-Leibler) divergence from the true posterior \cite{blei_ptm_2012, blei_review_2017, mani_variational_2020}. To this end, stochastic gradient descent or score function gradient estimation \cite{ranganath_bb_2014, paisley_variational_2012}
is used to maximize the ELBO (Evidence Lower Bound) objective \cite{braun_elbo_2010}, which is equivalent to minimizing the aforementioned KL divergence. One form \cite{hoffman_elbo_2016} of the ELBO objective $\mathcal{L}$, seen in equation \ref{eq_elbo}, depicts the trade-off between the expected complete log-likelihood, which encourages the variational distribution, q parameterized by $\nu$, to place its mass on the MAP (Maximum A Posteriori) estimate, and the negative entropy, which encourages q to be diffuse.

\begin{equation}
\label{eq_elbo}
\mathcal{L}(\nu) = \mathbb{E}_{q}[log(p(T,A,\theta))] - \mathbb{E}_{q}[log(q(\theta;\nu))]
\end{equation}

The framework we employed is called the Box's Loop \cite{blei_box_2014}, which describes an iterative process to develop latent variable models: model formulation, data analysis, performance evaluation, revision and repetition. This framework suits our task well as we posit a model with certain assumptions about the structure of the latent variables and use those to explain the data, evaluate the performance, and revise the model iteratively. 
\subsection{The Inference Task}
Following the approximation of the variational distribution, once we were satisfied with the obtained model, we proceeded to the application step of the Box's loop \cite{blei_box_2014}. In this step, the likelihood function from equation \ref{eq_posterior} was used to generate the predictions. Using the final estimates of the latent variables, we drew samples from $P(T_{pq}|u_p, v_q, bu_p, bv_q, \kappa_t, \psi_t, H)$ and $P(A_{pq}|u_p, v_q, bu_p, bv_q, \kappa_a,$ $\psi_a, H)$ to estimate the transaction score and the temporal affinity score, respectively. We then summed up the scores obtained to get a final score for the user $p$ and the product category $q$ under consideration.

\section{Experiments}
\subsection{Datasets}
We demonstrate our results on two real-world datasets. The first one is the well-known Instacart Orders dataset \cite{instacart_dataset}, which is a dataset of 3 million Instacart orders from over 200,000 users. Along with each user's basket, there is information about the sequence of products purchased in each order, the week and hour of day the order was placed, and a relative measure of time between orders. The transaction metadata contained a lot of other information on the products, aisles, and departments. The dataset was already partitioned into training and test sets. Since we are looking at a category affinity problem and since the dataset did not have any explicit categories, some preprocessing was required. We generated Word2Vec embeddings of the product names in the train set using the Continuous Bag-of-Words model \cite{mikolov_word_2013} and performed k-means clustering on the vectors thus obtained. We experimented with different values of the number of clusters, different ways to assign the initial cluster centers etc., and evaluated the quality of the clustering using visual quality assurance, as well as cluster purity analysis using aisles as labels. We then adopted the clusters as product categories. \par
The second dataset is a private dataset from a large-scale e-Commerce company. We collected six months’ worth of grocery transaction data, which resulted in a much denser interaction matrix as compared to the Instacart dataset. We used the first five months as our training set and the last month as the test set. From the transaction metadata, the relevant columns, such as the user id, the product category, the transaction date, and the event epoch (exact epoch at which the transaction happened), were extracted.  Since we are trying to understand category affinities of users, it was important that the users had a certain minimum number of data points. So, we decided to filter out users who had less than a given threshold number of transactions, thus retaining only the engaged users, as both the baseline models and our model would not be able to make informed predictions for the low-engagement users. This is the classic cold-start problem in recommender systems \cite{bobadilla_coldstart_2012, rashid_preferences_2008, lika_coldstart_2014}.  A related issue is the long-tail problem in e-Commerce and online grocery \cite{richards_longtail_2018, singer_longtail_2012, bailey_longtail_2008, kilkki_practical_2007}. Both of these are caused by a variety of factors, including but not limited to sellers introducing new items and even product categories to capture the dynamically changing user interests,  seasonality effects, etc. Thus, we removed the product categories not present in both the training and the test sets as well. We also know from experience that there are a few large clients, such as grocery retailers, that buy in bulk. Since these would not be good representatives of individual buyers, we also filtered out users with more than an upper threshold number of transactions.\par
These preprocessing steps were performed on both the datasets, and this resulted in 135,000 users and 11,000 product categories for the Instacart dataset, and 180,000 users and 3,100 product categories for the private dataset. \par 

\subsection{Baselines}
To compare the performance of our model, we evaluated it against standard baselines from literature such as ItemPop (Pop) and implicit Matrix Factorization (MF), as well as advanced ones like Bayesian Personalized Ranking (BPR) as is standard practice \cite{wan_monotonic_2018, rendle_bpr_2009}. In addition to this, we also compared it with the plain Metapath2Vec model (M2V) \cite{dong_m2v_2017} without the additional introduction of temporal category affinity and the subsequent optimization using variational inference over the two matrices. 

The ItemPop model recommends categories to users based on the general popularity of the categories. It observes the interaction during the training period and simply takes that as the preference score. The second baseline is the Matrix Factorization model based on latent factors. An Alternating Least Squares optimization is done on the user-category transaction matrix and once the user and category embeddings are obtained, an inner product between the two is used to come up with the predicted user-category score. The third method for comparison was the BPR setup which is a pairwise ranking algorithm which has proved to be better than the standard matrix factorization techniques. Finally, we also used the embeddings obtained from the M2V model for recommendation and then compared the results of our model against these baselines.

\subsection{Implementation}
To implement Metapath2Vec, we formed a heterogeneous interaction graph from the datasets and traversed it using random walks of different lengths to get the context vectors for each node. Since parameterizing the lengths of the walks and the number of walks per node is a substantial problem in its own right, we needed a fast implementation of walks. This task was parallelizable and there was considerable improvement in performance through parallelization of walks for each node. Once the walks were done, we used the code \cite{m2v_code} provided by the authors of \cite{dong_m2v_2017} to implement the heterogeneous skip-gram model and obtain the Metapath2Vec embeddings. 

For the baselines ItemPop and Matrix Factorization, we used the popular recommender systems library of Apple, Turi Create \cite{tc_apple}. Turi Create has efficient models and modularized implementations for tasks like recommendations, object detection, image classification, image similarity, etc. For Bayesian Personalized Ranking, we implemented the baseline in Python using the model presented in \cite{rendle_bpr_2009}.

To perform black-box variational inference \cite{ranganath_bb_2014} for our model, we wrote a custom training loop \cite{mani_variational_2020} using Edward2 \cite{tran_edward_2016, tran_deep_2017, tran_simple_2018}. Edward2 is a probabilistic programming language that provides core utilities in the NumPy and TensorFlow ecosystems to write models as probabilistic programs and manipulate a model's computation for flexible training and inference. Since both the posterior probability and the gradient could be computed in a distributed fashion, we leveraged TensorFlow to significantly speed up the training process. The training loop is detailed in algorithm \ref{vi_edw}. First, the log-likelihood is computed by sampling from the proxy posterior. Then, Edward2's tracing is used to record the model's computations for automatic differentiation. The KL divergence between the variational distribution and the prior distribution is computed and tracers are applied to write the ELBO. Different optimizers, learning schedules, and hyperparameter settings were tried.  
\begin{algorithm}
\scriptsize
  \caption{Variational Inference Training Loop}\label{vi_edw}
  \begin{flushleft}
        \textbf{INPUT:} Batch from Transaction matrix $T_b$, batch from Temporal Affinity matrix $A_b$, Transaction matrix $T$, Temporal Affinity matrix $A$, set of hyperparameters $H$, set of latent variables $\theta$, set of prior variables $\{u, v, bu, bv, \kappa_t, \psi_t, \kappa_a, \psi_a\}$
\end{flushleft}
  \begin{algorithmic}[1]
    \Procedure{Custom training loop}{$T_b$, $A_b$}
    
    \State variational\_family, trainable\_parameters $\gets$ Build variational distribution
    \State $qu, qv, qbu, qbv, q\kappa_t, q\psi_t, q\kappa_a, q\psi_a$ $\gets$ Sample posterior variables from the variational\_family
    %\State ${PP}_T, {PP}_L$ $\gets$ Set prior variables to the sample posterior values and obtain the posterior predictive distributions $P(T|\theta, H)$ and $P(L|\theta, H)$ using the corresponding terms from equation \ref{eq3}
    \State ${PP}_T, {PP}_A$ $\gets$ Obtain posterior predictive functions, $P(T|\theta, H)$ and $P(A|\theta, H)$, from equation \ref{eq_posterior} by setting prior variables to the sample posterior values
    \State $LL_{T_b}$, $LL_{A_b}$ $\gets$ Compute the log likelihood of $T_b$ and $A_b$ from ${PP}_T$ and ${PP}_A$, respectively
    \State Initialize $KL$ $\gets$ 0
    \For {prior\_variable, variational\_variable in [$(u, qu)$, $(v, qv)$, $(bu, qbu)$,
                $(bv, qbv),$ $(\kappa_t, q\kappa_t)$, $(\psi_t, q\psi_t)$,$(\kappa_a, q\kappa_a)$, $(\psi_a, q\psi_a)$]}
        \State $KL$ $\gets$ $KL$ + KL divergence between the distributions of the variational\_variable and the prior\_variable
        \EndFor 
    \State ELBO $\gets$ Compute ELBO using $KL$, $LL_{T_b}$, and $LL_{A_b}$ by rewriting equation \ref{eq_elbo} as  $\mathcal{L}(\nu) = \mathbb{E}_{q}[log(p(T,A|\theta))] - KL(q(\theta;\nu) \parallel p(\theta))$, where $p(\theta)$ is the prior over the latent vectors
    \State Loss $\gets$ -ELBO
    \State Get the gradients using the loss and the trainable\_parameters obtained
    \State Update the parameter values and use them in step 2 of the next iteration

    \EndProcedure
  \end{algorithmic}
\end{algorithm}

\subsection{Evaluation Metrics}

For each user, the ground truth was the list of product categories they engaged with in the test window. Here, by engagement, we mean that they should have bought items belonging to that product category. There are other ways to measure engagement, in terms of clicks, add-to-carts, etc., but transaction is a stronger, albeit sparser, signal. The predictions returned by the model were a list of product categories, ordered by the probability that the user would engage with items in that category in the test window. We compared our model with the baselines on the following well-known metrics \cite{manning_ir_2008, jarvelin_ir_2017, lioma_evaluation_2017}:

a) NDCG@k (Normalized Discounted Cumulative Gain): DCG measures the usefulness of a result based on its position in the ranking list \cite{wang_theoretical_2013}. The premise is that highly relevant entries are more useful when they appear earlier in a predictions list, and greater the relevance, greater the usefulness. The relevance for a product category is 1 if it appears in the top k predictions for a user and is present in the ground truth, and 0 otherwise. Ideal DCG (IDCG) is used to normalize this score to account for the varying lengths of the lists returned for different users. In the following formulae, $rel_i$ denotes the relevance of the entry at the $i^{th}$ position in the predictions list returned by the models.
\begin{equation}
DCG_k = \sum_{i=1}^{k} \frac{rel_i}{log_2(i+1)},  NDCG_k = \frac{DCG_k}{IDCG_k}\end{equation}\par 
\vspace{2mm}
b) Hit rate@k: Also known as recall or sensitivity, this measures the fraction of the relevant results that are retrieved in the top k predictions. A result is relevant for a user if the user has purchased at least one item from that product category in the test window.
\begin{equation}
\text{Hit rate} = \frac{\text{|(relevant results) $\cap$ (retrieved results)|}}{\text{|relevant results|}}\end{equation}\par 
\vspace{2mm}
c) MRR@k (Mean Reciprocal Rank): The reciprocal rank for a user is the inverse of the position of the first product category in the list returned by the model that is present in the ground truth for that user \cite{radev_evaluating_2002}. We consider only the first k entries in the returned list and average the reciprocal ranks over all the queries, i.e. users. Even though this metric only cares about the highest-ranked relevant result, along with MAP (described next), this gives us a good signal about the user experience given our ranking model. In the following formula, $pos_i$ represents the position of the first prediction for the $i^{th}$ user such that user $i$ has bought an item from that product category in the test window. 
\begin{equation}
MRR = \frac{1}{|U|} \sum_{i=1}^{|U|} \frac{1}{pos_i}\end{equation}\par
\vspace{2mm}
d) MAP@k (Mean Average Precision): Almost equivalent to the PR-AUC (Area Under the Precision-Recall Curve), the average precision is the area under the curve obtained by plotting the precision and recall at every position in a ranked list of predictions \cite{zhu_recall_2004}. It is the average of the precision obtained every time a new positive sample is recalled. This is a very useful metric when there is class imbalance and we care more about the positive class, as this metric is very sensitive to the improvements for the positive class. Mean of the average precision scores over a set of queries i.e. users, gives the MAP. In the following formulae, $P(i)$ is the precision at position $i$, $\Delta r(i)$ is the change in recall from position $i-1$ to $i$, \#rel is the total number of relevant product categories for that user (up to k), and $|U|$ is the number of users in the test set. 
\begin{equation}
AP = \sum_{i=1}^{k}P(i){\displaystyle \Delta} r(i) = \frac{\sum_{i=1}^{k}P(i)rel_i}{\text{\#rel}}, MAP = \frac{\sum_{j=1}^{|U|} AP_j}{|U|}\end{equation}\par 
\vspace{2mm}
e) LAUC@k (Limited Area Under the Curve) : A potential drawback of ROC (Receiver Operating Characteristic) curves \cite{fawcett_roc_2006} in recommender systems is that AUC is equally affected by swaps at the top or the bottom of the returned list, but users care more about the results at the top of the list. Limited AUC \cite{schroder_goals_2011} partly mitigates this issue by calculating the area under the part of the curve formed by the top k recommendations. It assumes that all the other relevant recommendations are distributed uniformly over the rest of the ranking list until all entries are retrieved. Even though a swap at the top or the bottom of the top-k list has the same effect on the LAUC value, swaps below the top k don't affect the AUC. Also, a top-k list that contains more relevant entries will yield a higher AUC score. \par 
\vspace{2mm}

\section{Results \& Analysis}
\subsection{Quantitative Results}
Table \ref{instacart_metrics_table} presents the results of our experiments on the Instacart dataset while Table \ref{wmt_metrics_table} illustrates the results on the private e-Commerce dataset. The first thing to note is that the results are of different orders. The reason for that is the different number of categories in the two datasets as well as the difference in the density of transactions. The Instacart dataset's categories were from the clusters formed during the k-means clustering step which had a different granularity and precision as compared to the category classification in the private e-Commerce dataset (which has human curated, more accurate product category classification). 

For the Instacart data, there is approximately a 20-30\% lift in NDCG for the top 5 recommendations from the baselines. ItemPop (Pop) lacks personalization and that clearly affects the results while plain Matrix Factorization (MF) is a notch below Bayesian Personalized Ranking (BPR) and Metapath2Vec (M2V) embeddings. BPR and M2V roughly perform to similar levels with the former outperforming the latter  by a slight margin. Introducing the second matrix clearly has a pronounced effect on the higher spots for our model (VI). As for the overall pool of recommendations, the hit rate increasing by roughly 30-60\% shows that the quality of the recommendation pool has gone up. The Mean Reciprocal Rank going up by roughly 20\% indicates that the first relevant recommendation is showing up in higher slots. 

For the private e-Commerce dataset, there is approximately a 40\% lift in NDCG with respect to the closest baseline. ItemPop is well below the other baselines while MF is closer to BPR for this dataset. The reason could be that the interactions are denser in the private e-Commerce dataset than in the Instacart data. This could also be the reason for the plain Metapath2Vec algorithm clearly outperforming the Bayesian Personalized Ranking in the top slots as the denser interaction graph provides richer contexts for highly connected nodes. As we go to the bottom slots, Metapath2Vec seems to be getting slightly worse than BPR.  This illustrates that the Metapath2Vec algorithm has some difficulties in category granularity which will be further discussed in Section \ref{case_study}. The VI model does maintain a healthy advantage over BPR even in the lower slots. For a business with billions of dollars in revenue, the increase in metrics seen here translates to several thousand items being viewed due to the increased category level traffic and as a result, several million dollars in revenue. 

\begin{table}[t]

%\centering
\footnotesize{
\begin{tabular}
{
%|p{1.1cm}|p{0.5cm}|p{0.8cm}|p{0.8cm}|p{0.8cm}|p{0.8cm}|p{0.8cm}|
|C{1.1cm}|C{0.5cm}|C{0.8cm}|C{0.8cm}|C{0.8cm}|C{0.8cm}|C{0.8cm}|
}\hline
%\hline
%\backslashbox{Method}{Metric}
%&\makebox[3em]{Rank}&\makebox[3em]{Pop+EE}&\makebox[3em]{LMF+}
%&\makebox[3em]{6/3}&\makebox[3em]{6/4}\\\hline\hline
\diagbox[width=5.4em, height=7.5em]{\textbf{Metric}}{\textbf{Method}} & & \textbf{\newline \newline Pop} & \textbf{\newline \newline MF} &\textbf{\newline \newline BPR} & \textbf{\newline \newline M2V} & \textbf{\newline \newline VI} \\  
%\multirow{2}{*}{}\footnotesize{\textbf{Method}} & & \textbf{\newline Pop \newline + EE} & \textbf{\newline LMF + EE} &\textbf{\newline Pop} & \textbf{\newline LMF} & \textbf{\newline VI Model} \\ 
%\textbf{\footnotesize{Metric}}&\ &\ &\ &\ &\ &\\
 \hline
\multirow{6}{*}{\textbf{NDCG}}&@5 &0.025 &0.040  &0.053 &0.050 & \textbf{0.061}\\ 
&@10 &0.020 &0.033  &0.039 &0.039 &\textbf{0.052}\\
&@15 &0.018 &0.032  & 0.037 &0.036 & \textbf{0.047}\\
&@20 &0.018 &0.032  &0.036 & 0.035 & \textbf{0.046}\\
\hline
\multirow{6}{*}{\textbf{HR}}&@5 &0.023 &0.042  &0.057 &0.057 & \textbf{0.073}\\ 
&@10 &0.014 &0.028  &0.035 &0.035 & \textbf{0.054}\\
&@15 &0.012 &0.026  &0.029 &0.028 & \textbf{0.043}\\
&@20 &0.011 &0.027  &0.027 &0.026 & \textbf{0.040}\\
\hline
\multirow{6}{*}{\textbf{MRR}}&@5 &0.051 &0.076  &0.089 &0.088 & \textbf{0.102}\\ 
&@10 &0.051 &0.077  &0.089 &0.088 & \textbf{0.107}\\
&@15 &0.052 &0.079  &0.089 &0.088 & \textbf{0.105}\\
&@20 &0.052 &0.080  &0.089 &0.088 & \textbf{0.105}\\
\hline
\multirow{6}{*}{\textbf{MAP}}&@5 &0.012 &0.020  &0.024 &0.024 & \textbf{0.030}\\ 
&@10 &0.009 &0.014  &0.017 &0.017 & \textbf{0.022}\\
&@15 &0.008 &0.013  &0.016 &0.015 & \textbf{0.020}\\
&@20 &0.008 &0.013 & 0.015 & 0.015 & \textbf{0.019}\\
\hline
\multirow{6}{*}{\textbf{LAUC}}&@5 &0.512 &0.522  &0.529 &0.529 & \textbf{0.542}\\ 
&@10 &0.509 &0.516  &0.520 &0.520 & \textbf{0.529}\\
&@15 &0.508 &0.516  &0.518 &0.518 & \textbf{0.526}\\
&@20 &0.508 &0.516  &0.518 &0.517 & \textbf{0.525}\\
\hline

\end{tabular}
}

\captionof{table}{Comparison of evaluation metrics across models on the Instacart orders dataset}\label{instacart_metrics_table}
\end{table}

\begin{table}[t]

%\centering
\footnotesize{
\begin{tabular}{
%|p{1.1cm}|p{0.5cm}|p{0.8cm}|p{0.8cm}|p{0.8cm}|p{0.8cm}|p{0.8cm}|
|C{1.1cm}|C{0.5cm}|C{0.8cm}|C{0.8cm}|C{0.8cm}|C{0.8cm}|C{0.8cm}|
}\hline
%\hline
%\backslashbox{Method}{Metric}
%&\makebox[3em]{Rank}&\makebox[3em]{Pop+EE}&\makebox[3em]{LMF+}
%&\makebox[3em]{6/3}&\makebox[3em]{6/4}\\\hline\hline
\diagbox[width=5.4em, height=7.5em]{\textbf{Metric}}{\textbf{Method}} & & \textbf{\newline \newline Pop} & \textbf{\newline \newline MF} &\textbf{\newline \newline BPR} & \textbf{\newline \newline M2V} & \textbf{\newline \newline VI} \\  
%\multirow{2}{*}{}\footnotesize{\textbf{Method}} & & \textbf{\newline Pop \newline + EE} & \textbf{\newline LMF + EE} &\textbf{\newline Pop} & \textbf{\newline LMF} & \textbf{\newline VI Model} \\ 
%\textbf{\footnotesize{Metric}}&\ &\ &\ &\ &\ &\\
 \hline
\multirow{6}{*}{\textbf{NDCG}}&@5 &0.191 &0.256  &0.271 &0.336 & \textbf{0.462}\\ 
&@10 &0.124 &0.186  &0.206 &0.219 & \textbf{0.305}\\
&@15 &0.097 &0.157  & 0.178 &0.171 & \textbf{0.241}\\
&@20 &0.082 &0.140  &0.161 & 0.145 & \textbf{0.207}\\
\hline

\multirow{6}{*}{\textbf{HR}}&@5 &0.179 &0.240  &0.259 &0.350 &\textbf{0.519}\\ 
&@10 &0.090 &0.149  &0.173 &0.176 &\textbf{0.267}\\
&@15 &0.061 &0.117  &0.142 &0.119 &\textbf{0.185}\\
&@20 &0.046 &0.102  &0.126 &0.091 &\textbf{0.146}\\
\hline

\multirow{6}{*}{\textbf{MRR}}&@5 &0.348 &0.404  &0.408 &0.480 &\textbf{0.482}\\ 
&@10 &0.348 &0.406  &0.411 &0.480 &\textbf{0.482}\\
&@15 &0.348 &0.407  &0.412 &0.480 &\textbf{0.482}\\
&@20 &0.348 &0.407  &0.412 &0.480 & \textbf{0.482}\\
\hline

\multirow{6}{*}{\textbf{MAP}}&@5 &0.089 &0.142  &0.154 &0.198 &\textbf{0.321}\\ 
&@10 &0.045 &0.080  &0.092 &0.099 &\textbf{0.164}\\
&@15 &0.030 &0.050  &0.068 &0.067 &\textbf{0.112}\\
&@20 &0.023 &0.046 & 0.056 & 0.052 & \textbf{0.087}\\
\hline

\multirow{6}{*}{\textbf{LAUC}}&@5 &0.589 &0.619  &0.629 &0.641 &\textbf{0.751}\\ 
&@10 &0.543 &0.574  &0.585 &0.586 & \textbf{0.632}\\
&@15 &0.528 &0.556  &0.568 &0.557 & \textbf{0.591}\\
&@20 &0.520 &0.548  &0.559 &0.543 & \textbf{0.570}\\
\hline

\end{tabular}
}

\captionof{table}{Comparison of evaluation metrics across models on the private e-Commerce dataset}\label{wmt_metrics_table}
\end{table}

\subsection{Qualitative Case-Study} 
In this subsection, we present plots of three different subspaces of the embedding space where our variational inference based algorithm improves upon the Metapath2Vec algorithm, in terms of capturing the semantics of product clusters. These plots are from the private dataset, which has accurate product category and product family labels. Product families are collections of similar product categories in the catalog. The vector plots have been projected down to two dimensions. The vectors are also normalized by rows to get the two embeddings on the same scale. Different shapes have been used to indicate the different algorithms while the different colors indicate the different product families.

\begin{figure}
\begin{center}
\resizebox{250pt}{!}{\includegraphics{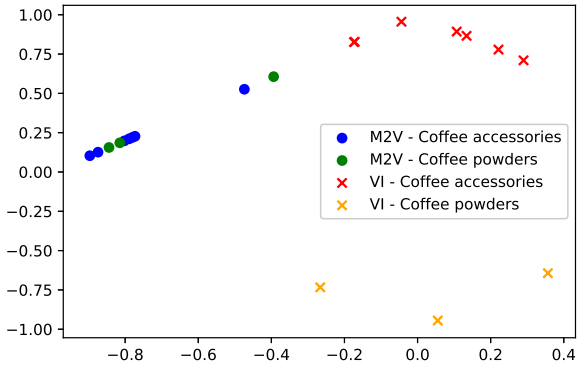}}
%{\includegraphics[scale=0.3]{graphical_model_vi_2.png}}
\caption{Embeddings corresponding to the following product families (and product categories): Coffee accessories (Decanters, Filters, Grinders, Percolators, Drippers, Pods), and Coffee powders (Whole Bean, Instant Coffee, Ground Coffee)}
\label{fig:coffee_family}
\end{center}
\end{figure}

Figure \ref{fig:coffee_family} presents a case study of categories related to coffee. While the Metapath2Vec algorithm does a good job of clustering together relevant categories, there is no real distinction between the coffee accessories family and the coffee powders family. VI, initialized with embeddings from Metapath2Vec, on the other hand is able to clearly separate out the coffee powders from the coffee accessories. One of the reasons for this could be the temporal affinity signals: people are likely to buy coffee powders way more often than they buy coffee accessories. 

\begin{figure}
\begin{center}
\resizebox{250pt}{!}{\includegraphics{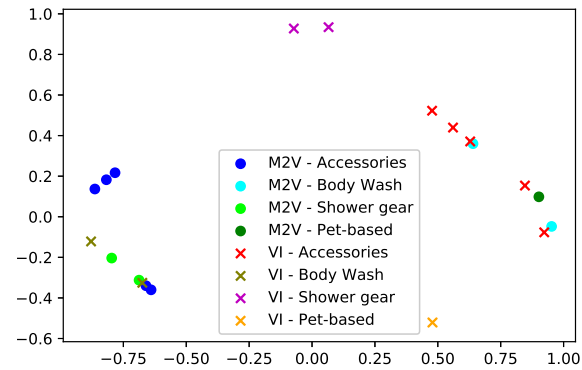}}
%{\includegraphics[scale=0.3]{graphical_model_vi_2.png}}
\caption{Embeddings corresponding to the following product families (and product categories): Shower accessories (Organizers, Stalls \& Bases, Curtains \& Liners, Curtain Rods, Curtain Rings, Bath Stools \& Benches), Body wash (Shower Gels, In-shower Lotions), Shower gear (Portable Shower Gear, Shower Caps), and Pet-based shower products (Pet Shower \& Bath Accessories)}
\label{fig: shower_family}
\end{center}
\end{figure}

\begin{figure}
\begin{center}
\resizebox{250pt}{!}{\includegraphics{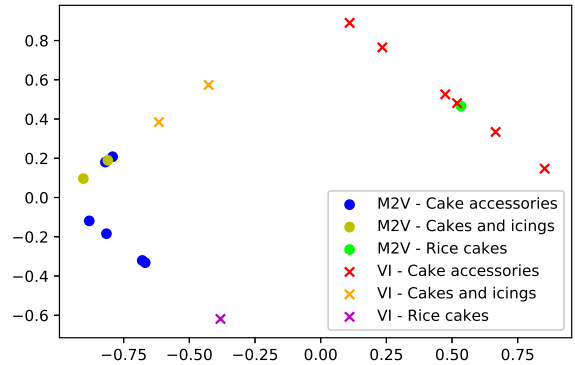}}
%{\includegraphics[scale=0.3]{graphical_model_vi_2.png}}
\caption{Embeddings corresponding to the following product families (and product categories): Cake accessories (Boards \& Circles, Servers, Turntables, Boxes, Pans, Stands), Cakes and icings (Cakes, Cake Toppers), and Rice cakes (Rice Cakes)}
\label{fig:cake_family}
\end{center}
\end{figure}

Similar to this, Figure \ref{fig: shower_family} presents a case study in categories related to shower items. There are four families here: shower accessories, body wash, shower gear, and pet-based shower products. VI again provides a clear distinction between the four product families. This can again be attributed to the temporal affinity signal. Figure \ref{fig:cake_family} has results on cakes. To illustrate the general grasp of semantics in both models, we have presented figure 3, where both algorithms are clearly able to distinguish rice cakes from normal cake based families. VI does a better job in separating out cake accessories from cakes and icings. 

\label{case_study}

\section{Conclusion \& Future Work}

In this work, we capture users' temporally changing category affinities by leveraging their repeat purchase patterns to develop scalable user and category embeddings. We also exploit the graphical structure in the user-category interaction data using Metapath2Vec to initialize the embeddings. We then cast this as an optimization problem and solve it iteratively by employing variational inference. These scalable representations thus obtained then help us tackle the category recommendation problem. We demonstrate the effectiveness of the representations learnt both quantitatively, by showing that our model outperforms the baselines on two real-world datasets, and qualitatively, by showing the semantics learnt by the embeddings for different product categories.

In terms of future work, one could try other initialization strategies to capture the semantics from the graphical structure of the user-category data. One could also extend our work and employ it in conjunction with works that leverage category experts \cite{hwang_experts_2016} for efficient recommendations. Another direction to explore would be to develop other ways to capture users' temporally changing category affinities and to employ domain knowledge to model the different distributions.

\bibliographystyle{ACM-Reference-Format}
\bibliography{sample-base}

\appendix

\end{document}